\documentclass[11pt]{article}

\begin{document}

\thispagestyle{empty}

\begin{center}
\vspace*{2cm}
{\LARGE $R^0$ Cosmology ?}
\vskip14mm
{\bf Gerald Vones}
\vskip2mm
Energy Department, Federal Ministry of Economics and Labour\\
Schwarzenbergplatz 1, A-1014 Wien, Austria\\
gerald.vones@bmwa.gv.at
\vskip17mm
\begin{abstract}
It is postulated that the action of the FRW-universe is the cosmological 
term of Einstein's theory (no curvature term - ``$R^0$ Cosmology''). The
expansion  equation emerging from the embedding of this most simple brane world
with variable speed  of light is deduced. The universal dimensionless coupling
constant of gravity is addressed. Some implications on the deep problems of
cosmology are discussed.
\end{abstract}
\end{center}
\newpage
\setcounter{equation}{0}
\setcounter{page}{1}

\section{Introduction}

Recent observations have brought new life into the discussion on the
cosmological models. For about two decades  the standard theoretical idea
has been an early period of inflationary expansion \cite{Guth,Linde},
amongst these  models being such without big bang \cite{Chauvet} or pre
big bang evolution \cite{Turner}. But doubts have  increased and
alternatives have been constructed, in particular theories with variable
speed of light (VSL)  \cite{Mo,Albrecht}. One of the most recent
suggestions is the so called ekpyrotic universe \cite{Steinhardt}.

In this letter I present an alternative which might be able to teach
something even if it failed to describe our  universe, since it is a very
special one: It is the most simple and transparent model ever possible;
the only relevant parameter which is open for tuning is the number of (large)
spacelike dimensions. It is based on the  suspicion that the
Einstein-Hilbert action principle applied to the
Friedmann-Robertson-Walker (FRW)  universe does not contain too few terms,
but too many. By this I mean that first only the cosmological  term has to
be taken into account as a source (similar to the de Sitter model; for the
further I will address  this as the vacuum as usual) and that second the
curvature term is absent.  Although my considerations have been
distinct from string and brane theory, the outcome can be
addressed as a brane  world; indeed the action is nothing but those given
by Pav\v si\v c \cite{Pav} with the ``matter density''  set equal to a
constant (set $\omega=-\Lambda/G$ in his equation (1) and understand the
intrinsic metric to  be those of the bare 4-manifold). This trivialization
of the ansatz does of course  not address all problems of cosmology, but
it does address the most deep ones in an adequate manner, exactly as it
should be the case for the FRW approximation.

As a matter of fact, General Relativity Theory is a highly accurate description of gravitation for 
noncosmological situations. A central part of my considerations was dedicated to the question whether
this fact is in conflict with the postulate underlying the theory presented in this paper. 
I cannot give the final answer, but have gained
some optimism that the unification will be possible in a satisfactory manner. For the 
time being one should be aware of the actual character of the FRW universe: An idealization, which 
inhibits most of the gravitational dynamics of the cosmic fluid by assumption. My ansatz gives a justification 
for this idealization in a highly transparent manner.
 
The paper is organized as follows: In section 2 the action principle is
formulated and the geometric  meaning of the concepts involved is
clarified. In section 3 the expansion equation is derived. In  section 4
the behaviour of standard clocks is discussed and the integral conservation
law for the  Hamiltonian is derived. In section 5 the alternative of a
3-torus universe is introduced. In section 6  the idealized cosmic fluid
is taken into account as additional source and the universal coupling
constant  of gravity is addressed. In section 7 some implications on the
most deep problems of cosmology  are discussed.

\section{Action Principle and Embedding of the FRW Universe}

I postulate that the cosmos is described by the cosmological term, that means 
the action of the universe idealized as FRW model is its 4-volume exclusively:
\begin{equation}
S=-\frac{\Lambda}{G} \int\,dV_4\ \;,
\end{equation}
where $\Lambda$ is the cosmological constant and $G$ is Newton's
constant. The negative sign expresses the fact that under the assumption
of spatial homogeneity and isotropy (see comments in section \ref{concl})
the variation leads to a maximum of the absolute value of the action. All
quantities  involved are understood to be real positive numbers.

I call this theory  ``$R^0$ Cosmology'' to express the fact that the curvature
scalar does not appear in the action.

The concept of extra dimensions comes in immediately in an ancient familiar
way, since the action  presented leads to a meaningful evolution equation only
if the curved universe is generated by an  embedding mechanism. The number of
dimensions of the embedding space is open (see also  chapter \ref{chtorus}). As
the simplest possibility I copy the embedding of the de Sitter universe  by
postulating the existence of a 5-dimensional quasi-Euclidean embedding space
with metric $diag (1,-1,-1,-1,-1)$.

The line element in the embedding space can be written as:
\begin{equation}
ds^2_{emb} = dt^2 - dR^2 -R^2[d\varphi^2 + sin^2 \varphi (d\vartheta^2 + sin^2 \vartheta d\chi^2)]
\end{equation}
with obvious meaning of the coordinates.

The expansion equation introduces one restriction, leading to an embedded 4-dimensional universe. Under the 
idealization of spatial homogeneity and isotropy the expansion equation only involves $t$ and $R$, thus the 
line element of the embedded FRW-universe is:
\begin{equation}
ds^2_{FRW}= (1-\dot{R}^2) dt^2 - R^2[d\varphi^2 
+ sin^2 \varphi (d\vartheta^2 + sin^2 \vartheta d\chi^2)]\;,
\end{equation}
where the dot means $d/dt$.

A de Sitter universe ( = vacuum) is a manifold lying in the spacelike region w.r.t. the origin of the 
embedding space fulfilling the expansion equation $R^2 - t^2 = const > 0$, thus being a hyperboloid invariant 
under rotation of axes in the embedding space (de Sitter group $SO(4,1)$). In contrast, any universe fulfilling 
a different expansion equation is not invariant under the de Sitter group and does define a preferred timelike 
direction in the embedding space.

\section{The expansion equation; Symmetry Breaking}

The 4-volume of the FRW-universe reads in terms of the coordinates introduced above:
\begin{equation}
dV_4=2\pi^2R^3\sqrt{1-\dot{R}^2}dt \; .
\end{equation}

\begin{figure}[h]
\setlength{\unitlength}{1cm}

\begin{picture}(10,6)

\put(4,1){\vector(1,0){5}}
\put(9.2,1){t}

\put(4,1){\vector(0,1){3}}
\put(4,4.2){R}

\thicklines

\put(4,1){\line(1,1){4}}
\put(7.5,4.25){null cone}

\qbezier(1,4.16)(4,1.32)(7,4.16)
\put(1,2.9){de Sitter}

\qbezier(4,1)(5.5,2.5)(6,2.5)
\put(3.5,0.6){Big Bang}
\put(6.05,2.55){Terminator}

\qbezier[15](6,2.5)(6.5,2.5)(8,1)

\end{picture}

\begin{center}
\caption{The de Sitter- versus the $R^0$-universe (the recontracting branch is dotted; 
see comments in section \ref{concl})\label{unis}}
\end{center}
\end{figure}
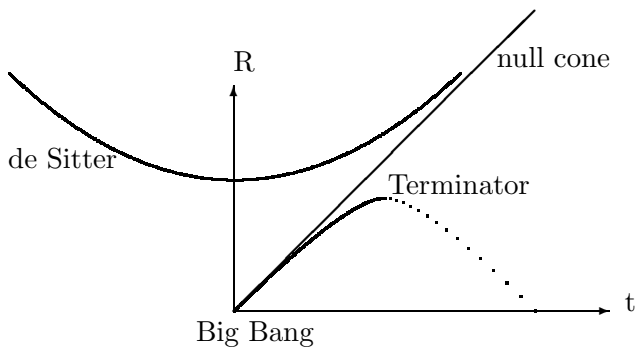

The variation w.r.t. $R$ and $dR/dt$ leads straightforwardly to the expansion equation:
\begin{equation}
\dot{R}^2=1-(\frac{R}{R_{max}})^6 \;,
\end{equation}
where the constant of integration is named in an obvious way, due to the fact
that the expansion  comes to a halt at a 3-sphere I name the ``terminator'';
there the natural boundary condition $dR/dt = 0$ is fulfilled. The second
integration leads to an elliptic integral. For the second constant of integration 
$R(0) = 0$ (big bang), consequently $dR(0)/dt =1$ is the natural choice. Since
the expansion equation is different from the de Sitter case the breaking of the
de Sitter group occurs and a preferred frame - the rest frame of the universe -
is defined; the associated center-of-mass worldline is timelike and lies within
the embedding space, but not within the universe.  In addition, the big bang
brakes the translation invariance of the embedding space and defines the 
unique origin which is a pointlike singularity of the universe, relative to
which the entire universe lies inside the future cone defined in the embedding
space. The situation is sketched in figure \ref{unis} in a qualitative manner.

%%%%%%%%%%%%%%%%%%%%%
%%%%%%%%%%%%%%%%%%%%%

\section{The role of time; Conservation law}

With regard to the time scale one can conclude that $t$ is already the quantity measured by the usual (atomic) clocks of observers at rest in the universe, since in terms of ``universal proper time'' $d\tau^2 = dt^2 - dR^2$ one arrives at the 
inequality $\frac{\tau}{R}\frac{dR}{d\tau} \leq \frac{1}{4}$, in conflict with the data on the age of the universe and the Hubble constant.

The consequence of this behaviour of clocks is a change over time of the velocity of light in the universe. But in contrast to the VSL-theories it starts with the value zero and increases to its maximum value at the terminator by virtue of (for a 
wave propagating along $d\vartheta = d\chi= 0$):
\begin{equation}
\frac{R d\varphi}{dt}=\sqrt{1-\dot{R}^2}=\frac{d\tau}{dt}
=(\frac{R}{R_{max}})^3:=c \;.
\label{velo}
\end{equation}
The term to the very left is the velocity of light measured with the help of
the atomic clocks.  It is proportional to the 3-volume of the universe; it is
normalized to unity in the embedding space, consequently to unity at the
terminator.

It is clear from equation (\ref{velo}) that in terms of ``universal proper time'' the velocity of 
light is constant; its natural unit is the value at the terminator. The expansion equation reads in terms of $\tau$:
\begin{equation}
(\frac{dR}{d\tau})^2=(\frac{R_{max}}{R})^6 - 1 \;.
\end{equation}

The variability of the velocity of light manifests itself also in the relation
between the fundamental constants. Here it is always understood that the
conversion of units is performed via the velocity of light in the embedding
space.

Since $R$ does not depend on the angles, the theory is one-dimensional and the
Lagrangian reads:
\begin{equation}
L_{vac}=-\frac{2\pi^2\Lambda}{G}R^3 \sqrt{1-\dot{R}^2}=
-\frac{2\pi^2\Lambda R_{max}^3}{G} c^2 := -Ec^2 \;,
\end{equation}
where the symbol $E$ is an abbreviation for the forefactor with the unit $mass$ or $energy$.

From this one arrives at total momentum (corresponding to the radial motion in the embedding space) and Hamiltonian:
\begin{eqnarray}
P_{vac}= \frac{\partial L_{vac}}{\partial \dot{R}}= 
\frac{2\pi^2\Lambda}{G}
\frac{R^3 \dot{R}}{\sqrt{1-\dot{R}^2}} = E \dot{R} \;,\\
H_{vac}=P_{vac}\dot{R}-L_{vac}=  
\frac{2\pi^2\Lambda}{G}
\frac{R^3}{\sqrt{1-\dot{R}^2}} = E  \;.
\end{eqnarray}
This means, the total energy is conserved as a consequence of the expansion equation and the energy density $E/(2\pi^2R^3)$ is - irrespective how ``relativistic'' the expansion of the universe may be - always inversely proportional to the 
3-volume.

\section{Alternative topology: 3-torus universe}
\label{chtorus}

Given the behaviour of the velocity of light discussed above the distance-redshift relation takes on the form:
\begin{equation}
z=\frac{H_0 D}{c_0}+\frac{H_0^2 D^2}{2c_0^2}(q_0+4)+\cdots 
\end{equation}
where $z$ is the redshift, $D$ is the distance of the object at the moment of measurement, $H_0$ is the Hubble 
constant at that moment, $c_0$ is the velocity of light at that moment and $q_0$ is the deceleration parameter 
derived from the Hubble constant. The difference to the redshift formula for constant velocity of light is the 4 in 
the parenthesis instead of a 1. This change per se makes the discrepancy with the new results on the expansion 
of the universe \cite{Riess} even stronger; but since the behaviour of the candles is influenced as well, only a 
deeper analysis can give a sufficient answer.

At this point I would like to mention another property of the model: Should there be overwhelming evidence 
that our universe is flat in the sense of zero intrinsic curvature of the spacelike sections at constant 
universal time then this fact can be taken account of by postulating a 7-dimensional quasi-Euclidean 
embedding space with one positive and 6 negative components of the metric.

In this case the line element can be written as:
\begin{equation}
ds^2_{emb} = dt^2 - dR_1^2 - dR_2^2 - dR_3^2 
-R_1^2 d\varphi^2 - R_2^2 d\vartheta^2
- R_3^2 d\chi^2
\end{equation}
again with obvious meaning of the coordinates.

The expansion equation has to introduce 3 restrictions simultaneously:
\begin{equation}
R_1=R_2=R_3:=\frac{R(t)}{\sqrt{3}}
\end{equation}
and the line element of the embedded universe becomes:
\begin{equation}
ds^2_{torus}= (1-\dot{R}^2) dt^2 - 
\frac{R^2}{3}(d\varphi^2 + d\vartheta^2 + d\chi^2) \;.
\end{equation}
In terms of $R$ the volume element differs from the expression for the 3-sphere universe only by an 
overall geometric factor, which is irrelevant for the expansion equation and can be regarded as 
incorporated in the cosmological constant. Thus the expansion equation and all the other conclusions 
remain unaffected; for the further I will not differentiate between these two topologies.

\section{Coupling of matter and radiation}

The real universe is filled with matter and radiation. As long as the spatial homogeneity and isotropy is kept as
an approximation the postulate underlying $R^0$ Cosmology states that neither the action nor the embedding mechanism
are affected. This means, the Lagrangian of the free particles must be equal to minus the coupling term as a tautology.
In the matter dominated era, when the energy density of the fluid emerges from the 
conserved total mass $M$ of the particles at rest in the universe, this can be
modeled by simple vector coupling. The free Lagrangian of the particles and the momentum vector 
derived from it are as in SRT in 5 dimensions:

\begin{equation}
L_{mat}+L_{coupl}=-M\sqrt{1-\dot{R}^2}+K\eta_{\mu\nu}P_{mat}^\mu P_{vac}^\nu 
\;,
\end{equation}
where the indices run over 5 components (only 2 of them being relevant) and $K$ is a coupling
constant with unit $mass^{-1}$.

This leads to the additional terms in the Lagrangian:
\begin{equation}
L_{mat}+L_{coupl}=M\sqrt{1-\dot{R}^2}(-1+KE) \;.
\label{Lag}
\end{equation}
To make the contributions vanish in total, the coupling constant must be equal to $E^{-1}$.

In the general case of a gas of particles with individual masses $M_n$ the coupling can be retained as above 
if the definition of the preferred frame is complemented by the condition that in this frame the 
total angular momentum around 
the $t$-axis (i.e. the total spatial momentum inside the universe) of the fluid is zero. The Hamiltonian 
relevant for the coupling term can then be derived from the free Lagrangian in terms of the 
``macroscopic'' quantities and does incorporate the pressure contribution. 
In equations (\ref{one}) and (\ref{two}) the 
subscript $mat$ is left away; furthermore only one spatial degree of freedom inside the universe is written
out explicitly:
\begin{eqnarray}
L&=&-\sum_nM_n\sqrt{1-\dot{R}^2-R^2\dot{\varphi}_n^2}\;,
%=-\sum_n\frac{M_n}{\sqrt{1-\dot{R}^2-R^2\dot{\varphi}_n^2}}
%+\sum_n\frac{M_n\dot{R}^2 }{\sqrt{1-\dot{R}^2-R^2\dot{\varphi}_n^2}}+
%+\sum_n\frac{M_nR^2\dot{\varphi}_n^2}{\sqrt{1-\dot{R}^2-R^2\dot{\varphi}_n^2}}
\label{one}\\
p_R&=&\frac{\partial{L}}{\partial{\dot{R}}}=
\sum_n\frac{M_n\dot{R}}{\sqrt{1-\dot{R}^2-R^2\dot{\varphi}_n^2}}\;,
\nonumber\\
p_{\varphi_n}&=&\frac{\partial{L}}{\partial{\dot{\varphi_n}}}=
\frac{M_nR^2\dot{\varphi}_n}{\sqrt{1-\dot{R}^2-R^2\dot{\varphi}_n^2}}
\;, \qquad \sum_np_{\varphi_n}=0\;,
\nonumber\\
H&:=&p_R\dot{R}-L\nonumber\\
&=&\sum_n\frac{M_n}{\sqrt{1-\dot{R}^2-R^2\dot{\varphi}_n^2}}
-\sum_n\frac{M_nR^2\dot{\varphi}_n^2}{\sqrt{1-\dot{R}^2-R^2\dot{\varphi}_n^2}}
\; .
\label{two}
\end{eqnarray}
This can be extended to massless particles by the usual change from the masses to the
frequencies $M_n\rightarrow h\nu_n\sqrt{1-\dot{R}^2-R^2\dot{\varphi}_n^2}=0$, from
 which both the free Lagrangian and the coupling term are null.

Taking account of the expansion equation, equation (\ref{Lag}) can be formulated in a 
somewhat different way. For this purpose I write the inverse coupling constant as:
\begin{equation}
K^{-1}=E:=M\alpha^2 \;,
\label{emc2}
\end{equation}
where $\alpha$ is a pure number; the square is chosen for convenience.
This yields:
\begin{equation}
H_{mat}+H_{coupl}=\frac{M}{\sqrt{1-\dot{R}^2}}-\alpha^{-2}\frac{R_{max}^3E}{R^3}=
c^{-1}(M-\alpha^{-2}E)  \;.
\label {attr}
\end{equation}
The formula says that the kinetic energy of the particles moving in the embedding space is the 
same as the energy per 3-volume of the vacuum up to the factor between $E$ and $M$.

The value of  $\alpha$ is unknown so far. But there are some observations which help to narrow the variety 
of possibilities and allow to address $\alpha$ as the universal dimensionless coupling constant of gravity. 
First, the theory contains a fundamental length different from the Planck length, second the ratio of these 
two fundamental lengths is the largest number achievable in such a way, third the equation (\ref{emc2}) is a natural 
candidate to contain such a fundamental constant and fourth the energy balance can be interpreted in terms
of gravitation. Due to equation (\ref{attr}) the particles gain potential energy during 
expansion by virtue of gravitational attraction; to the same amount they loose kinetic energy due to the 
deceleration of the radial motion, while the energy of the vacuum ( = FRW $R^0$-universe) itself is unaffected.

Up to factors of order of magnitude unity the most plausible values for the quantities involved are:
\begin{eqnarray}
\alpha^{-2}R^3_{max}\approx \alpha l^3_{pl}\quad
\qquad\qquad\mbox{i.e.}\qquad\qquad&\alpha \approx \frac{R_{max}}{l_{pl}}
\label{al}
\end{eqnarray}
\begin{eqnarray}
\Lambda \approx l^{-2}_{pl}\qquad\qquad&E\approx \alpha^3m_{pl}\qquad\qquad
&M\approx \alpha m_{pl}  \;.
\end{eqnarray}
These equations can be addressed as a modern version of Dirac's Large Number Hypothesis, stating 
that the total mass of the cosmic fluid is equal to the maximum radius of the universe if both are measured 
in Planck units. The relations furthermore introduce the dimensionless 
coupling constant as the factor between the integration constant and the (square root of the) 
cosmological constant; the high value of 
$\Lambda$ is possible, since it enters the Lagrangian only as an overall multiplicative factor.

Assuming that $R_{max}/R_{today}$ (and all the ratios derived from that, in particular the present 
value of the velocity of light) is not a large number but rather of order of magnitude unity one can convince 
oneself that equation (\ref{al}) makes sense by comparing the mass density $M/(2\pi^2R^3)$ - order of 
magnitude $m_{pl}/(l_{pl}R^2_{today})$ - to the measured matter density which is somewhere around 
$10^{-30} g/cm^3$ and the speculative value of $R_{today}$ which is somewhere around $10^{30} cm$.

\section{Discussion and conclusion}
\label{concl}

The theory presented includes two results which can - at least in principle - be tested directly 
by observation, that are the increase over time of the velocity of light and the expansion equation.

For a decision on the appropriateness, further analysis has to be performed. At the present stage the 
most impressive thing is the way in which the most deep problems of cosmology are addressed:

{\bf Existence problem}:
The universe is generated by breaking of the translation and rotation symmetry of the embedding 
space. But the unbroken symmetry - the quasi-Euclidean embedding space - remains visible in the
imagination, since it is necessary to allow a simple deduction of the expansion equation. Per definition,
physical mechanisms neither allow to penetrate the region ``before'' the big bang nor elsewhere outside 
the universe, which is a subset of the embedding space.

{\bf Flatness problem}:
No tuning of terms is necessary, since only one term survives in the action. The matter and radiation 
related terms sum up to zero.

{\bf Problem of the cosmological constant}:
The cosmological constant, present in the action as an overall multiplicative factor, does not appear 
in the expansion equation; any nonsingular value is possible.

{\bf Singularity problem}:
The big bang and the terminator are straightforwardly derived from the expansion equation. The initial 
singularity is directly connected to the breaking of the translation invariance of the embedding space. 
It defines a natural starting condition for the expansion; the infinite energy density is a simple consequence 
of the geometric singularity. The terminator is a branch point.
(As long as one stays with the FRW model the second branch can be added as usual and the universe let shrink 
again to a point. But having an eye on the entropy problem discussed below one should be cautious with 
any speculation regarding the evolution beyond the terminator.)

{\bf Horizon problem}:
The big bang is a pointlike singularity, thus the entire universe is causally connected at the beginning - 
irrespective of the fact that the velocity of light is zero at that moment. Furthermore, in the embedding 
space the entire universe lies inside the future cone emerging from the big bang.

{\bf Homogeneity Problem and Entropy problem}:
That the universe is homogeneous and isotropic at the beginning can be understood as a consequence 
of the symmetry of the embedding space. Thus the low entropy singularity \cite{penrose} occurs 
naturally. The increase of the information content over time is also adequately reflected, since the manifold 
is unstable as a consequence of the negative sign of the action; thus the universe aims to enlarge its 3-volume by 
making $R$ vary as a function of the angular coordinates. Although the model can of course not incorporate 
the full dynamics of the cosmic fluid, it does exhibit its most fundamental feature, that is the 
increase of entropy over time.\\

The most unsatisfactory aspect of the model is the lack of an inherent mechanism 
which fixes the number of large dimensions of spacetime to its actual value; more or less the same 
is the case for the embedding space. Although this situation is not really new, it is much in contrast 
to the transparency of the rest of the theory.

\vskip15mm \noindent 
{\bf Acknowledgments}: I thank Professors C.B. Lang and H. Mitter for discussions at an earlier stage of this work.


\begin{thebibliography}{12}
%
\bibitem{Guth}
A.H. Guth, Phys. Rev. D {\bf 23}, 347 (1981)
%
\bibitem{Linde}
%
A.D. Linde, Phys. Lett. B {\bf108}, 389 (1982)
%
\bibitem{Chauvet}
P. A. Chauvet and L. O. Pimentel, Phys. Rev. D {\bf 54}, 7158 (1996)
%
\bibitem{Turner}
M.S. Turner and E.J. Weinberg, Phys. Rev. D {\bf 56}, 4604 (1997)
%
\bibitem{Mo}
J.W. Mo, Int. J. Mod. Phys. {\bf D2}, 351 (1993)
%
\bibitem{Albrecht}
A. Albrecht and J. Magueijo, Phys. Rev. D {\bf 59}, 043516 (1999)
%
\bibitem{Steinhardt}
J. Khoury, B.A. Ovrut, P.J. Steinhardt, N. Turok, hep-th/0103239
%
\bibitem{Pav}
M. Pav\v si\v c, Phys. Lett. A {\bf116}, 1 (1986), republished as
gr-qc/0101075 
%
\bibitem{Riess}
A.V. Filippenko, A.G. Riess, astro-ph/0008057
%
\bibitem{penrose}
R. Penrose, Singularities and Time-Asymmetry, in : S.W. Hawking and W.
Israel  (Editors), {\bf General Relativity: An Einstein Centenary},
Cambridge University Press (1979)
%
\end{thebibliography}
\end{document}